\documentclass[aps,prl,reprint,superscriptaddress,citeautoscript,showpacs]{revtex4-1}

\usepackage{graphicx}
\usepackage{xspace}		
\usepackage[final]{showkeys}	

\newcommand{\Tc}{\ensuremath{T_c}\xspace}
\newcommand{\TN}{\ensuremath{T_N}\xspace}
\newcommand{\Jc}{\ensuremath{J_c}\xspace}
\newcommand{\Hc}{\ensuremath{H_{c2}}\xspace}
\newcommand{\Hpc}{\ensuremath{\vec H\parallel{}c}\xspace}
\newcommand{\Hpab}{\ensuremath{\vec H\parallel{}ab}\xspace}
\newcommand{\RE}{RENi$_2$B$_2$C\xspace}
\newcommand{\Tm}{TmNi$_2$B$_2$C\xspace}
\newcommand{\Er}{ErNi$_2$B$_2$C\xspace}

\newcommand{\Y}{YNi$_2$B$_2$C\xspace}

\newcommand{\unit}[1]{\ensuremath{~\mathrm{#1}}}
\newcommand{\unitnospace}[1]{\ensuremath{\mathrm{#1}}}
\newcommand{\K}{\unit{K}}
\newcommand{\T}{\unit{T}}

\newcommand{\uV}{\unit{\mu{}V}}
\newcommand{\mm}{\unit{mm}}
\newcommand{\um}{\unit{\mu{}m}}

\newcommand{\degree}{\unitnospace{^\circ}}

\renewcommand{\vec}[1]{\mathbf{#1}}

\newcommand{\etal}{\textit{et~al.\ }}	
\newcommand{\vs}{vs\xspace}						

\begin{document}

\title{Strong Enhancement of the Critical Current at the Antiferromagnetic Transition\\in ErNi$_2$B$_2$C Single Crystals}

\date{\today}

\author{M.~Weigand}
\author{L.~Civale}
\author{F.\,J.~Baca}
\author{Jeehoon~Kim}
\affiliation{Materials Physics and Applications Division, Los Alamos National Laboratory, Los Alamos, New Mexico 87545, USA}
\author{S.\,L.~Bud'ko}
\author{P.\,C.~Canfield}
\affiliation{Ames Laboratory, US DOE and Department of Physics and Astronomy, Iowa State University, Ames, Iowa 50011, USA}
\author{B.~Maiorov}
\affiliation{Materials Physics and Applications Division, Los Alamos National Laboratory, Los Alamos, New Mexico 87545, USA}

\begin{abstract}
We report on transport and magnetization measurements of the critical current density $J_c$ in ErNi$_2$B$_2$C single crystals that show strongly enhanced vortex pinning at the N\'{e}el temperature $T_N$ and low applied fields. The height of the observed $J_c$ peak decreases with increasing magnetic field in clear contrast with that of the peak effect found at the upper critical field. We also performed the first angular transport measurements of $J_c$ ever conducted on this compound.  They reveal the correlated nature of this pinning enhancement, which we attribute to the formation of antiphase boundaries at $T_N$.
\end{abstract}


\maketitle



Single quanta of magnetic flux enter a type II superconductor in the form of vortices when it is exposed to a magnetic field $H$. This is a big setback for applications, because when an electrical current $J$ is applied, vortices move and energy is dissipated. This movement can be arrested by non-superconducting defects that pin vortices by lowering the system energy. The interaction between vortices and pinning centers has been studied extensively for decades, focused mainly on the so-called vortex core pinning (caused by the local suppression of the superconducting order parameter) \cite{MacManus-Driscoll2004, Maiorov2009}. Less explored is the  interplay between vortices and magnetic media, which could offer pinning forces superior to those from core pinning \cite{Bulaevskii1985, Bulaevskii2000, Jan2003, Blamire2009}. A common difficulty of studying magnetic pinning is the inability of separating the magnetic and core contributions. It is, therefore, of great advantage to study systems where the magnetic transition takes place inside the superconducting phase, allowing one to compare the behavior with and without the magnetically ordered phase. 

There are a considerable number of materials in which to study the coexistence between superconductivity and ordered magnetic phases, such as the Chevrel-phases \cite{Chevrel1986}, CeCoIn$_5$ \cite{Petrovic2001}, and recently iron pnictides \cite{Kamihara2008, Rotter2008, Cruz2008}. However, the rare earth-nickel-borocarbide family (\RE, where RE is a rare earth element) \cite{Nagarajan1994, Cava1994, Siegrist1994} has several advantages, namely a relatively high superconducting transition temperature \Tc and the tunability of the ratio between its magnetic and superconducting ordering temperatures, which can be changed by using different rare earth elements \cite{Canfield1998, Budko2006}. The readily available high-purity single crystals allow one to study the effects of magnetic pinning without any significant defect (non-magnetic) contribution \cite{Canfield1998}.

Among the borocarbide family the compound with RE = Er has a \Tc of $10.5\K$ and a N\'{e}el temperature \TN of $6.0\K$ \cite{Cho1995}. The occurrence of antiferromagnetism directly influences the superconducting properties, as seen in the upper critical field \vs temperature curve, where $\Hc(T)$ is slightly suppressed just below $T = \TN$ for \Hpc and has an inflection point for \Hpab \cite{Cho1995}. This is a consequence of the local moments ordering in the antiferromagnetic phase, as was shown experimentally \cite{Cho1995, Budko2000} and theoretically \cite{Machida1980, Ramakrishnan1981}.

In addition to the occurrence of antiferromagnetism, at $T = \TN$ a tetragonal to orthorhombic crystal structure transition takes place \cite{Detlefs1997}. The resulting twin boundaries were shown by Bitter decoration, magneto-optical, and scanning Hall-probe experiments to act as pinning centers at lower temperatures ($T < \TN$) \cite{Saha2000, Vinnikov2005}. The pinning mechanism was proposed to be caused by a ferromagnetic spin component parallel to the crystallographic $c$ axis localized at the twin boundaries \cite{Saha2000}.

Although borocarbides (and particularly the compound with RE = Er) have been extensively studied, few critical current measurements by electrical transport have been performed, with the exception of the seminal work by Gammel \etal \cite{Gammel2000}, exploring the weak ferromagnetism below $T = 2.3\K$ \cite{Canfield1996, Kawano1999, Choi2001, Kawano-Furukawa2002}. This was shown to lead to an enhanced critical current density \Jc, which was ascribed to an increase in pinning force due to pair breaking by the ferromagnetism \cite{Gammel2000}. Canfield \etal \cite{Canfield2001b} pointed out that the data in Ref.~\cite{Gammel2000} extrapolated to $\Jc = 0$ at \TN, indicating a clear linkage between pinning and the antiferromagnetic order. However, this extrapolation could not be confirmed because transport experiments were only done for $T < 4.2\K$ to take advantage of the cooling power of liquid helium \cite{Gammel2000}.

The lack of transport experiments is to a great extent due to heating problems caused by the applied current being very high because of the large cross section of single crystals. To the best of our knowledge, the present work is the first report both of transport measurements on \Er single crystals at temperatures around \TN and of angular \Jc measurements performed on this compound.

In this Letter, we present the results of transport and magnetization measurements on \Er single crystals, which revealed a local maximum in $\Jc(T)$ at $T = \TN$. We study this large increase in \Jc as a function of field strength and orientation, and we find that it occurs only for fields oriented along the crystallographic $c$ axis. Unlike the peak effect near \Hc, the height of the newly discovered maximum decreases with increasing magnetic field. We rule out pinning from twin boundaries as we are able to determine its angular fingerprint at lower temperatures, which differs from the increase in \Jc observed at \TN. We attribute the findings to vortex pinning due to antiphase boundaries between antiferromagnetic domains.



Large, homogeneous single crystals of \Er were grown using the Ni$_2$B flux growth technique \cite{Canfield2001}. The samples used for transport measurements were polished mechanically down to a thickness of $40-60\um$ parallel to the $c$ axis before being cut to pieces  $\sim\!\!250\um$ wide and $\sim\!\!1.5\mm$ long. Sputtered Au contacts allowed us to apply currents along the $ab$-planes while measuring the voltage across the sample. We observed a slight reduction of \Tc ($\sim\!\!0.5\K$) due to sample preparation, which is consistent with the previously reported increase in \Tc after annealing \cite{Miao2002}. The small decrease in \Tc does not affect the overall results of the present study; it left both \TN and the shape of $\Hc(T)$ unchanged.

Transport measurements were performed in three systems characterized by different cooling methods: (i) a Quantum Design, Inc.\ PPMS with a semi-closed He system, (ii) a variable temperature insert with flowing He, and (iii) with the sample immersed in liquid He. In all systems we used high-precision rotators and the maximum Lorentz force configuration  ($\vec{J} \perp \vec{H}$). The critical current density was obtained from $IV$-curves using a $1\uV$ criterion. Measurements of $R(T)$ and $R(H)$ gave the upper critical field \vs temperature, using a value of $R$ corresponding to $90\%$ of the normal conducting resistance. The comparison of the \Jc results obtained in the three systems allowed us to assess the magnitude of heating effects.

Magnetization measurements were carried out in a Quantum Design, Inc.\ MPMS Superconducting Quantum Interference Device (SQUID). \Jc was obtained using the Bean critical state model with $\Jc = \frac{|m_i|}{a b c} \cdot \frac{4}{b(1-b/3a)}$, where $a \geq b$, $m_i$ is the irreversible magnetic moment, and $a$, $b$, and $c$ are the sample length, width, and thickness, respectively \cite{Bean1962}.



In Fig.~\ref{fig:PPMS_0+30deg+Jc(T,HIIc)_contour_plot}(a) we present a \Jc transport measurement \vs temperature at $\mu_0H = 0.5\T$ for \Hpc. Starting at low temperature, $\Jc(T)$ decreases rapidly as $T$ increases, and a rough extrapolation of $\Jc(T)$ leads to $0$ at $T = \TN$, the same behavior found in Ref.~\cite{Gammel2000}. Our new measurements presented here, however, show that \Jc instead changes abruptly at \TN. The slope of $\Jc(T)$ is much lower at $T > \TN$ than below, and a peak in $\Jc(T)$ is observed at \TN, corresponding to a two-fold increase compared to the value just below and above it. These are the key results of this Letter. Finally, at a temperature just above $7\K$ we observe the well-documented peak-effect at \Hc.

\begin{figure}[htbp]
	\centering
	\includegraphics[width=0.65\columnwidth]{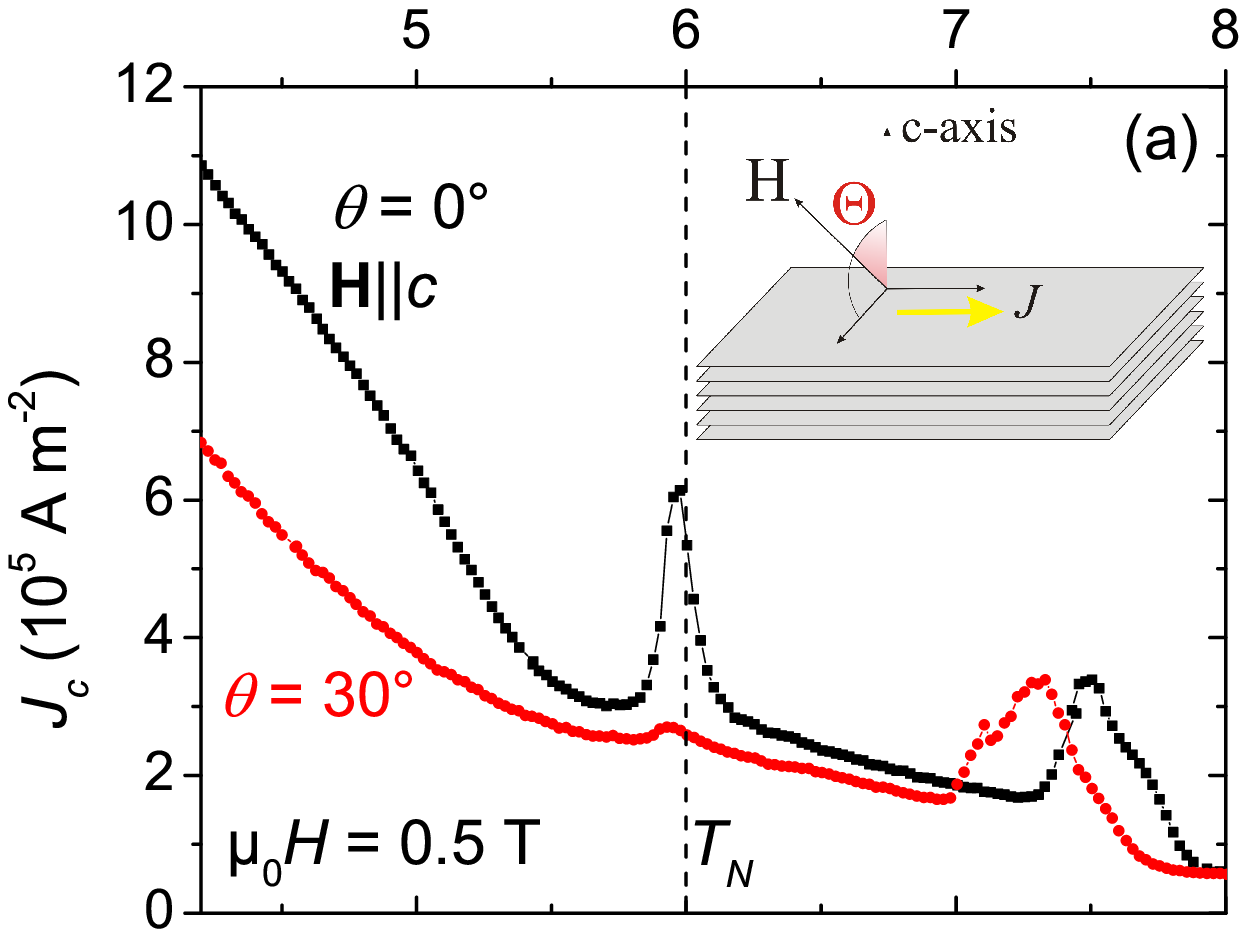}\\
	\includegraphics[width=0.65\columnwidth]{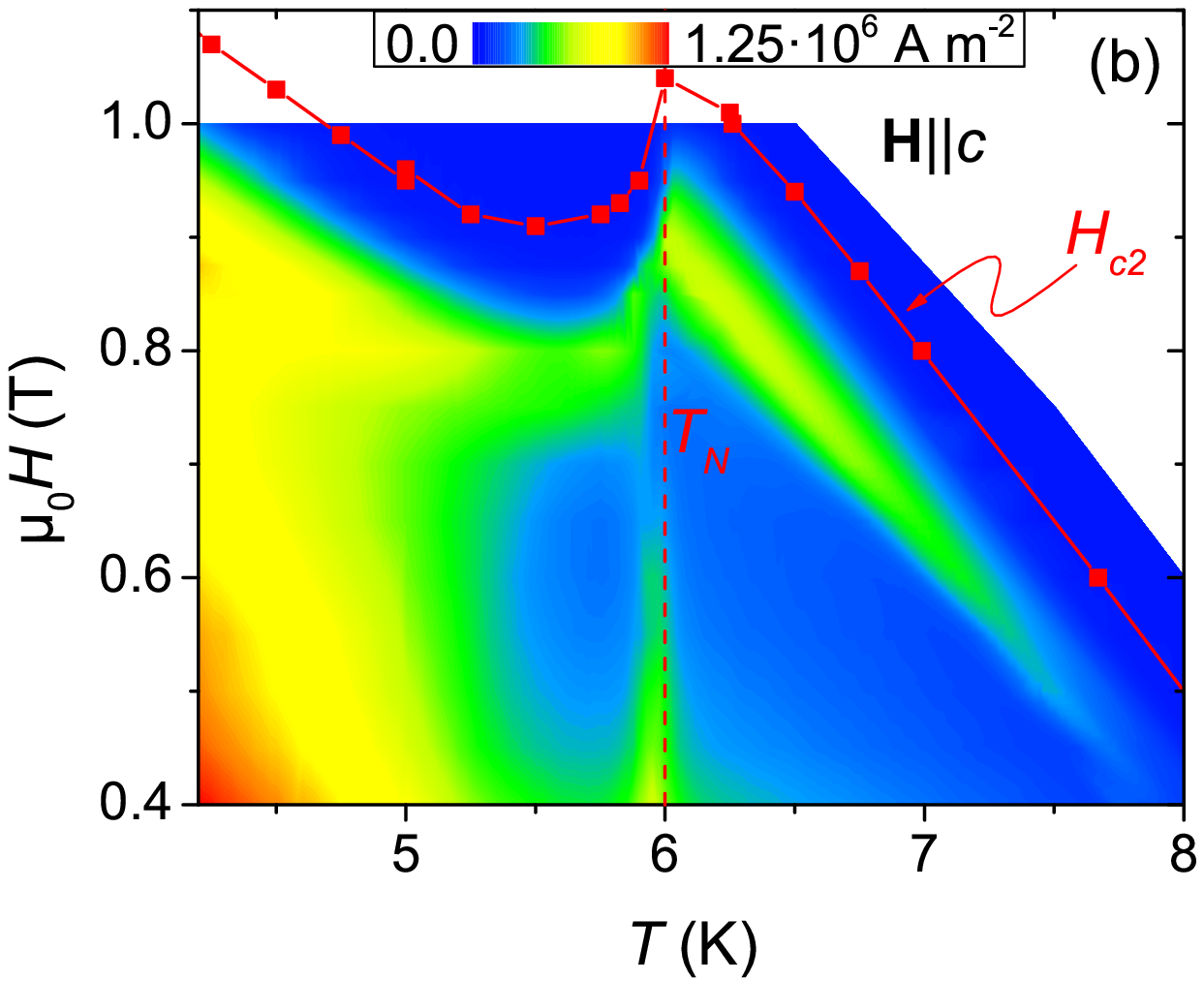}
	\caption{\label{fig:PPMS_0+30deg+Jc(T,HIIc)_contour_plot}(Color online) (a) Transport measurements of \Jc for \Hpc and $\theta = 30\degree$. A new peak in \Jc is observed at $T = \TN$ for \Hpc. The sketch gives the measurement geometry. (b) Contour plot of \Jc as a function of temperature and magnetic field \Hpc showing the field range where the peak at $T = \TN$ is present.}
\end{figure}

The maximum at $T = \TN$ occurs over a range of fields, as seen in the contour plot in Fig.~\ref{fig:PPMS_0+30deg+Jc(T,HIIc)_contour_plot}(b), obtained from multiple $\Jc(T)$ measurements at different \Hpc, where different colors represent different values of \Jc. Note that the height of the \Jc peak at \TN shows the opposite field dependence than that of the one observed near \Hc, indicating a different origin of the two maxima. Whereas the peak near \Hc becomes more pronounced as $H$ is increased, the peak at \TN decreases with increasing $H$ until it has essentially vanished at $\mu_0H = 0.7\T$. This leads us to speculate that the peak at \TN has a magnetic origin, as magnetic pinning tends to be attenuated as the magnetic field modulation of the vortices becomes smaller. Our observations are consistent with the dip found in the dynamic magnetic susceptibility of \Er single crystals at $T = \TN$, measured by Prozorov \etal \cite{Prozorov2009} using a tunnel-diode resonator technique. They explained their results by a pinning enhancement due to the occurrence of antiferromagnetic order, accompanied by large magnetic fluctuations. The decrease in the peak height with increasing $H$ also rules out the drastic change in $\Hc(T)$ at \TN as the origin for the maximum in $\Jc(T)$, since the peak disappears far below $\Hc(\TN)$. The absence of a $\Jc(T)$ peak at $\theta = 30\degree$, shown in Fig.~\ref{fig:PPMS_0+30deg+Jc(T,HIIc)_contour_plot}(a), is further proof that this peak is not associated with the sudden change in $\Hc(T)$ at \TN, since for $\theta = 30\degree$ this feature in $\Hc(T)$ is still visible (not shown).

It is clear from the \Jc$(T)$ measurements at $\theta = 0\degree$ and $30\degree$ that this effect is strongly angle-dependent. Thus, further insight can be gained from the dependence of \Jc on field orientation, since it allows one to distinguish between the superconducting and the magnetic anisotropy.  

\begin{figure}[htbp]
	\centering
	\includegraphics[width=0.65\columnwidth]{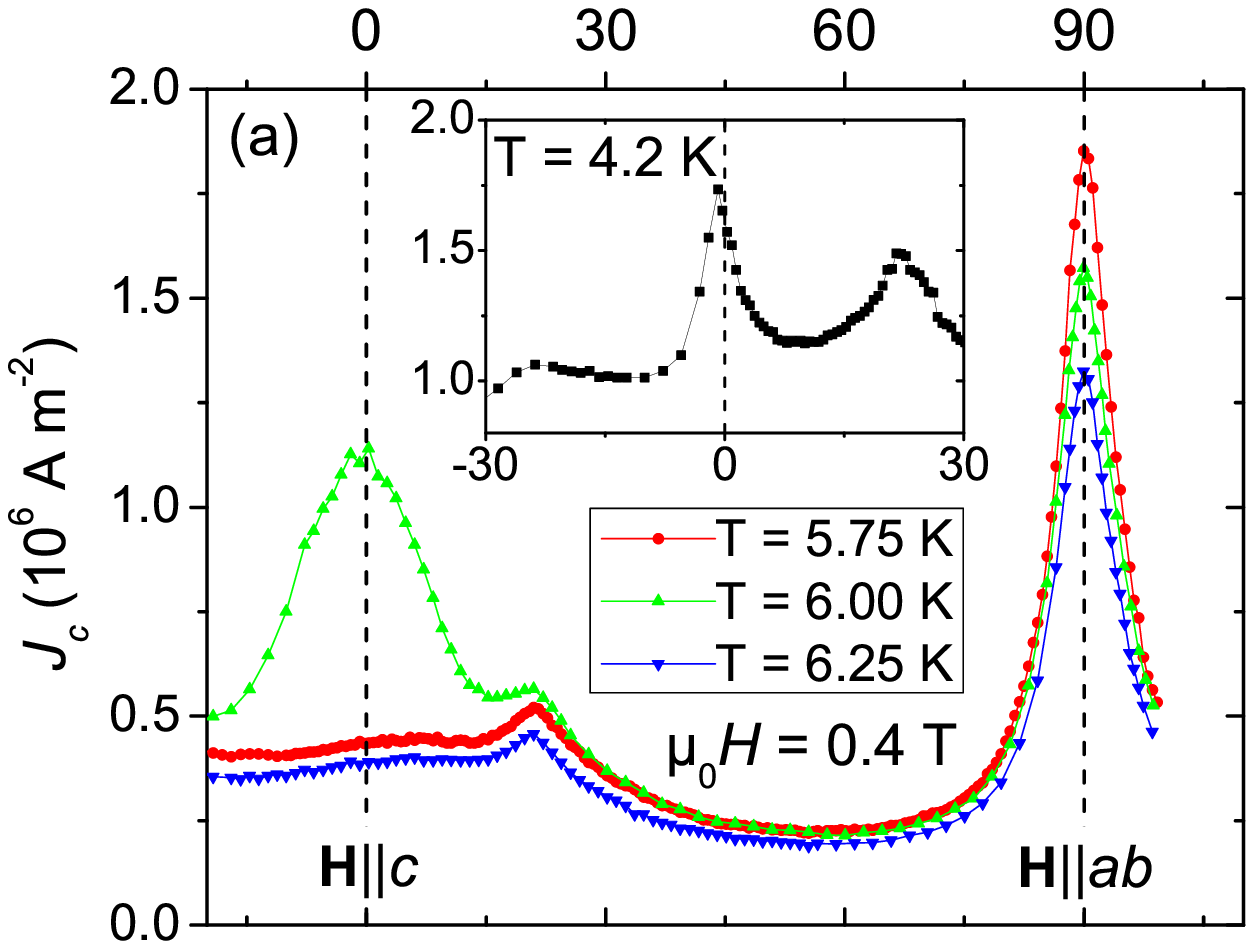}\\
	\includegraphics[width=0.65\columnwidth]{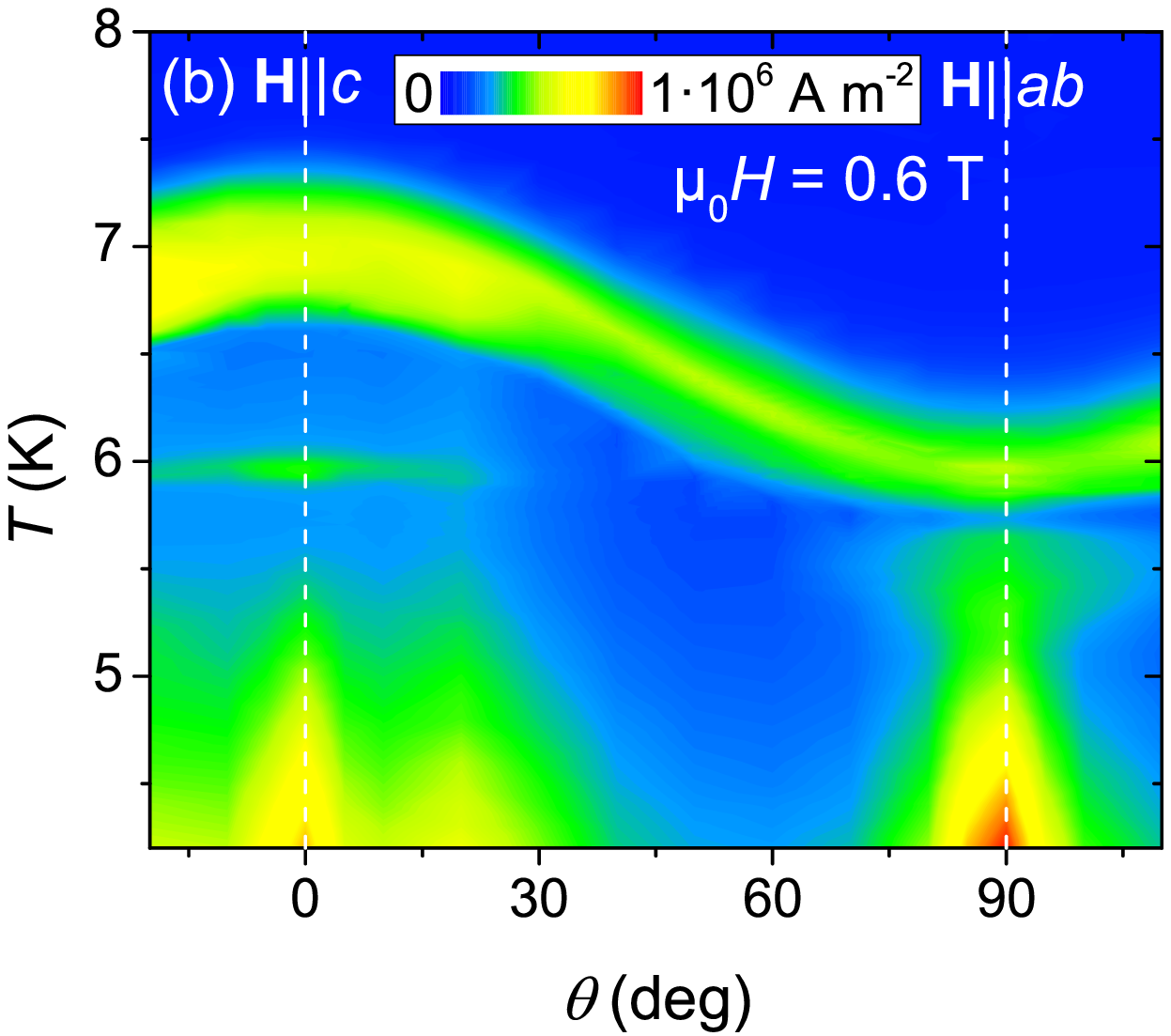}
	\caption{\label{fig:MLRot+PPMS_angular}(Color online) (a) Angular transport \Jc measurements at $\mu_0H = 0.4\T$ reveal the narrow temperature range at \TN over which the peak at \Hpc occurs. (b) The contour plot of \Jc as a function of magnetic field orientation $\theta$ and $T$ shows the location of the peak at \TN and $\theta = 0\degree$ in the $T$--$\theta$ space.}
\end{figure}

The obtained angular \Jc curves are reproduced in Fig.~\ref{fig:MLRot+PPMS_angular}(a). They reveal rich and complex features. Despite the small change of the absolute value of \Hc with field orientation, we found that \Jc has a pronounced maximum for \Hpab at all temperatures investigated.  This indicates strong pinning for \Hpab, in contrast to what would be expected from the anisotropy of \Hc, where $\Hc^{\parallel{}c} > \Hc^{\parallel{}ab}$. It also differs from what has been observed in non-magnetic \Y thin films, where both \Jc and \Hc show the same anisotropy \cite{Wimbush2003, Wimbush2004}. The maximum we find in $\Jc(\theta)$ around \Hpab can be explained by magnetic pinning, and it is consistent with the angular dependence of the magnetic susceptibility of the Er ions, namely $\chi \sim 0$ for \Hpc, while for \Hpab the full Curie-Weiss form is observed \cite{Budko2006}. We also find small maxima at $\theta = \pm22\degree$ both above and below \TN, which are not related to the sample shape and will require further investigation.

At $T = \TN$ a wide peak centered at \Hpc is visible, whereas at temperatures only $0.25\K$ above and below a plateau is observed. The peak in the angular dependence of \Jc indicates that the nature of this pinning is correlated. Naturally one is tempted to ascribe this maximum to the already reported pinning by a ferromagnetic spin component localized at magnetic twin boundaries \cite{Saha2000}; however we have measured a much narrower $\Jc(\theta)$ peak at lower temperatures [$T = 4.2\K$, see the inset of Fig.~\ref{fig:MLRot+PPMS_angular}(a)], which is clearly due to the presence of twin boundaries, as has been observed in Bitter decoration and scanning Hall-probe experiments \cite{Saha2000, Vinnikov2005}. Its different shape indicates that the \Jc peak at \TN is \emph{not} caused by twin boundaries. The peak due to twin boundaries appears for $T < \TN$ and exhibits a monotonic growth as the temperature decreases. The height of this peak can be roughly extracted when comparing $\Jc(T)$ at $\theta = 0\degree$ and $30\degree$ shown in Fig.~\ref{fig:PPMS_0+30deg+Jc(T,HIIc)_contour_plot}(a). The disappearance of this peak just below \TN is also confirmed by the angular dependence at $T = 5.75\K$ [see Fig.~\ref{fig:MLRot+PPMS_angular}(a)]. The temperature evolution of $\Jc(\theta)$ indicates that the peak at \TN has a different origin than that at low temperatures, although both are correlated along the $c$ axis.

The dependence of the \TN peak on temperature and field orientation becomes even clearer in the contour plot in Fig.~\ref{fig:MLRot+PPMS_angular}(b). It can be seen that \Jc is enhanced only over a narrow temperature and angular range near $T = 6\K$ and $\theta = 0\degree$.

The correlated nature of the pinning and its location around \TN lead us to conclude that it is due to dynamic magnetic domains that are created at \TN and annihilated as the temperature decreases further. Domain walls are known to produce planar pinning potentials in ferromagnetic/superconducting hybrids \cite{Vlasko-Vlasov2008, Vlasko-Vlasov2012}. This poses the interesting question why the pinning enhancement occurs for \Hpc, while the easy axis of the magnetic moment for \Er is the $b$ axis \cite{Zarestky1995, Sinha1995}. This could be due to the magnetic moment being flipped to the $c$ axis; it has to be noted, however, that this is highly unlikely because of the high crystalline electric field anisotropy of the local moment sublattice. Another explanation could be vortices bending into the $ab$ planes at the magnetic domain boundaries in order to gain the Zeeman energy. As the vortices move away from the $c$ axis the energy gain gets gradually smaller, until at the $ab$ planes there is no gain since vortices have zero net gain. A phenomenological model indicates that $J_c\propto \cos^3(\theta)$ \cite{Lin2013tobepublished}. This gives a broader maximum than the one observed experimentally, nevertheless it can explain qualitatively the enhancement observed. The presented results and the proposed model motivate measurements in other compounds such as \Tm, which has its easy axis parallel to the $c$ axis \cite{Cho1995b}. It is worth noting that associated with the tetragonal to orthorhombic transition in \Er \cite{Detlefs1997}, strain occurs that could also be responsible for enhanced pinning \cite{Prozorov2012privcomm}. Whether this is indeed the case could be answered by investigating a borocarbide which does not undergo a structural transition at $T = \TN$, namely again \Tm.

In order to further explore the effects at lower fields, which are not accessible to transport measurements due to heating, different samples were taken from the same batch and investigated in the SQUID. This also serves as verification of our transport measurement results. After creating a Bean profile \cite{Bean1962} by applying a  field \Hpc larger than $2H^*$ (the field above which screening currents flow in the entire sample volume) at $T = 4.5\K$, the magnetic moment $m$ was measured \vs temperature at $\mu_0H = 0.02\T$. In order to capture the increase in \Jc around $T_N$ the Bean profile was generated at different temperatures near $T_N$ before setting $\mu_0H = 0.02\T$ and measuring $m(T)$.

We then compiled an $m(T)$ curve from the different branches, selecting the branch with the highest (absolute) value of $m$ for each temperature. The obtained curve  represents an envelope of all $m(T)$ branches. With the Bean model the irreversible magnetic moment then gives the critical current density \cite{Bean1962}.

\begin{figure}[htbp]
	\centering
	\includegraphics[width=0.8\columnwidth]{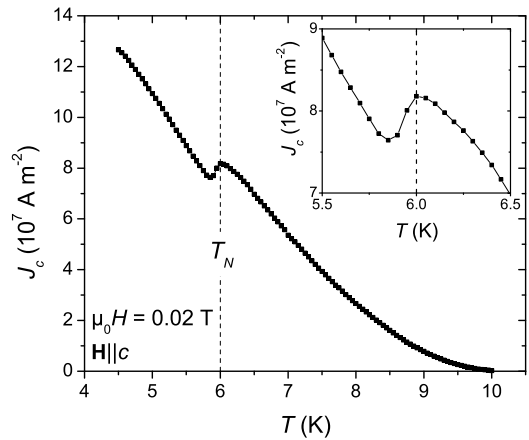}
	\caption{\label{fig:MPMS}The critical current density \vs temperature at $\mu_0H = 0.02\T$ (with \Hpc), obtained from magnetization measurements, exhibits a local maximum at $T = \TN$.}
\end{figure}

As can be seen in Fig.~\ref{fig:MPMS}, the obtained $\Jc(T)$ shows a peak exactly at $T = \TN$, similar to the local maximum observed in the transport measurements [see Fig.~\ref{fig:PPMS_0+30deg+Jc(T,HIIc)_contour_plot}(a)]. Although the shape of $\Jc(T)$ at this field resembles that of $\Hc(T)$, it is not directly governed by $\Hc(T)$, as the height of the \Jc peak obtained from magnetization curves decreases with increasing magnetic field, vanishing far below $\Hc(\TN)$, as in transport measurements. The peak in \Jc at \TN and its field dependence are seen in magnetic measurements for both a sample cut and polished to similar dimensions as the one used in transport experiments and a pristine single crystal with no treatment at all. This reinforces that the phenomenon is not sample dependent and that it can be experimentally verified by different techniques.

In summary, we have observed in transport and magnetization measurements an enhancement of the critical current density in \Er single crystals at the N\'{e}el temperature. By performing the first angular transport measurements we determined that this increase occurs for magnetic fields applied parallel to the $c$ axis, consistent with vortex pinning due to antiphase boundaries between antiferromagnetic domains. This study opens a new avenue to investigate the interaction between superconductivity and different magnetic phases, such as those known in Er, Ho and Tm borocarbides. In addition, new developments can be expected for other superconductors with coexistence of magnetic phases, such as underdoped copper- and iron-based superconductors \cite{Tranquada1995, Pratt2009}.


\begin{acknowledgments}
The authors are grateful to S.~Lin, C.\,D.~Batista, L.\,N.~Bulaevskii, and R.~Prozorov for useful discussions. This publication was made possible by funding from the Los Alamos LDRD Program, Project No.\ 20110138ER. The transport measurements were performed in part at the Center for Integrated Nanotechnologies and at the National High Magnetic Field Laboratory, both at Los Alamos National Laboratory. Work at Ames Laboratory (PCC and SLB) was supported by the U.S.\ Department of Energy, Office of Basic Energy Science, Division of Materials Sciences and Engineering. Ames Laboratory is operated for the U.S.\ Department of Energy by Iowa State University under Contract No.\ DE-AC02-07CH11358.
\end{acknowledgments}


\end{document}